\documentclass[12pt]{article}
\usepackage{times}
\usepackage{geometry}
\usepackage{graphicx}
\usepackage{sidecap}
\usepackage{times}
\usepackage{mathptmx}
\usepackage{wrapfig}
\usepackage{titlesec}
\usepackage[yyyymmdd]{datetime}
\usepackage[colorlinks=true, urlcolor=blue, citecolor=blue, linkcolor=blue]{hyperref}
\usepackage{ifthen}
\usepackage{afterpage}
\usepackage{appendix}
\usepackage{color}
\usepackage{cite}
\usepackage{lineno}
\usepackage{multirow}
\usepackage{parskip}
\usepackage[utf8]{inputenc}
\usepackage{enumitem,amssymb}
\usepackage{ragged2e}
\newlist{thematic}{itemize}{8}
\setlist[thematic]{label=$\square$}
\usepackage{pifont}
\newcommand{\Carnegie}{The Observatories of the Carnegie Institution for Science, 813 Santa Barbara St., Pasadena, CA 91101, USA}

\newcommand{\EPFL}{Institute of Physics, Laboratory of Astrophysics, Ecole Polytechnique Fédérale de Lausanne (EPFL), Observatoire de Sauverny, 1290 Versoix, Switzerland}

\newcommand{\EPFLEng}{STI IMT, École Polytechnique Fédérale de Lausanne (EPFL), 1015 Lausanne, Switzerland}
\newcommand{\FNAL}{Fermi National Accelerator Laboratory, Batavia, IL 60510}

\newcommand{\IRFU}{IRFU, CEA, Universit\'e Paris-Saclay, F-91191 Gif-sur-Yvette, France}

\newcommand{\KSU}{Kansas State University, Manhattan, KS 66506}

\newcommand{\LBL}{Lawrence Berkeley National Laboratory, Berkeley, CA 94720}

\newcommand{\MPHeidelberg}{Max Planck Institut für Astronomie, Königstuhl 17, D–69117 Heidelberg, Germany}

\newcommand{\NOAO}{National Optical Astronomy Observatory, 950 N. Cherry Ave., Tucson, AZ 85719 USA}

\newcommand{\NYU}{New York University, New York, NY 10003}

\newcommand{\OSU}{The Ohio State University, Columbus, OH 43212}
\newcommand{\OU}{Department of Physics and Astronomy, Ohio University, Clippinger Labs, Athens, OH 45701, USA}

\newcommand{\PI}{Perimeter Institute, Waterloo, Ontario N2L 2Y5, Canada}
\newcommand{\Pitt}{University of Pittsburgh and PITT PACC, Pittsburgh, PA 15260}

\newcommand{\Siena}{Siena College, 515 Loudon Road, Loudonville, NY 12211, USA}

\newcommand{\SLAC}{SLAC National Accelerator Laboratory, Menlo Park, CA 94025}

\newcommand{\Stanford}{Stanford University, Stanford, CA 94305}

\newcommand{\Tsinghua}{Department of Physics and Tsinghua Center for Astrophysics, Tsinghua University, Beijing 100084, P R China}

\newcommand{\UAS}{Department of Astronomy/Steward Observatory, University of Arizona, Tucson, AZ  85721}

\newcommand{\UCB}{Department of Astronomy, University of California Berkeley, Berkeley, CA 94720, USA}
\newcommand{\UCBP}{Department of Physics, University of California Berkeley, Berkeley, CA 94720, USA}
\newcommand{\UCBSSL}{Space Sciences Laboratory, University of California Berkeley, Berkeley, CA 94720, USA}

\newcommand{\UCL}{University College London, WC1E 6BT London, United Kingdom}

\newcommand{\UCSC}{University of California at Santa Cruz, Santa Cruz, CA 95064}

\newcommand{\UFL}{University of Florida, Gainesville, FL 32611}

\newcommand{\Umich}{University of Michigan, Ann Arbor, MI 48109}

\newcommand{\UNIPD}{Dipartimento di Fisica e Astronomia ``G. Galilei'',Universit\`a degli Studi di Padova, via Marzolo 8, I-35131, Padova, Italy}

\newcommand{\UR}{Department of Physics and Astronomy, University of Rochester, 500 Joseph C. Wilson Boulevard, Rochester, NY 14627, USA}

\newcommand{\Wyoming}{Department of Physics and Astronomy, University of Wyoming, Laramie, WY 82071, USA}
\newcommand{\Yale}{Department of Physics, Yale University, New Haven, CT 06520}
\newcommand{\Brandeis}{Department of Physics, Brandeis University, Waltham, MA 02453}
\usepackage{url}
\usepackage{soul}
\usepackage[dvipsnames]{xcolor}
\usepackage{amssymb, amsmath, epsfig}
\usepackage{hyperref}
\usepackage{tabularx}
\usepackage{subfigure}
\usepackage{epstopdf}
\usepackage{comment}
\usepackage{floatrow}
\newfloatcommand{capbtabbox}{table}[][\FBwidth]

\usepackage{blindtext}

\DeclareGraphicsExtensions{.ps}
\begin{document}

\raggedright
\huge
Astro2020 APC White Paper \linebreak

The MegaMapper: a $z > 2$ spectroscopic instrument for the study of Inflation and Dark Energy \linebreak
\normalsize

\noindent \textbf{Thematic Areas:} Ground Based Project, Cosmology and Fundamental Physics
 \linebreak

\textbf{Principal Authors:} \\ 
\ \\

Name: David J. Schlegel\\
Institution: Lawrence Berkeley National Laboratory \\
Email: djschlegel@lbl.gov 
 \linebreak
 
Name: Juna A. Kollmeier\\
Institution: Observatories of the Carnegie Institution of Washington \\
Email: jak@carnegiescience.edu
 \linebreak
 
\textbf{Proposing team:}\\
\ \\
Greg Aldering$^{1}$,
Stephen Bailey$^{1}$,
Charles Baltay$^{31}$,
Christopher Bebek$^{1}$,
Segev BenZvi$^{30}$,
Robert Besuner$^{3}$,
Guillermo Blanc$^{2}$,
Adam S. Bolton$^{20}$,
Mohamed Bouri$^{4}$,
David Brooks$^{24}$,
Elizabeth Buckley-Geer$^{29}$,
Zheng Cai$^{5}$,
Jeffrey Crane$^{2}$,
Arjun Dey$^{20}$,
Peter Doel$^{24}$,
Xiaohui Fan$^{6}$,
Simone Ferraro$^{1}$,
Andreu Font-Ribera$^{24}$,
Gaston Gutierrez$^{29}$,
Julien Guy$^{1}$, 
Henry Heetderks$^{3}$,
Dragan Huterer$^{11}$,
Leopoldo Infante$^{2}$,
Patrick Jelinsky$^{3}$,
Matthew Johns$^{2}$,
Dionysios Karagiannis$^{7}$,
Stephen M. Kent$^{29}$,
Alex G. Kim$^{1}$,
Jean-Paul Kneib$^{8}$,
Luzius Kronig$^{8}$,
Nick Konidaris$^{2}$,
Ofer Lahav$^{24}$,
Michael L. Lampton$^{3}$,
Dustin Lang$^{16}$,
Alexie Leauthaud$^{22}$,
Michele Liguori$^{7}$,
Eric V. Linder$^{3}$,
Christophe Magneville$^{9}$,
Paul Martini$^{19}$,
Mario Mateo$^{11}$,
Patrick McDonald$^{1}$,
Christopher J. Miller$^{11}$,
John Moustakas$^{27}$,
Adam D. Myers$^{21}$,
John Mulchaey$^{2}$,
Jeffrey A. Newman$^{14}$,
Peter E. Nugent$^{1}$,
Nathalie Palanque-Delabrouille$^{9}$,
Nikhil Padmanabhan$^{31}$,
Anthony L. Piro$^{2}$,
Claire Poppett$^{3}$,
Jason X. Prochaska$^{22}$,
Anthony R. Pullen$^{18}$,
David Rabinowitz$^{31}$,
Solange Ramirez$^{2}$,
Hans-Walter Rix$^{10}$,
Ashley J. Ross$^{19}$,
Lado Samushia$^{28}$,
Emmanuel Schaan$^{1}$,
Michael Schubnell$^{11}$,
Uros Seljak$^{1,12,13}$,
Hee-Jong Seo$^{17}$,
Stephen A. Shectman$^{2}$,
Joseph Silber$^{1}$,
Joshua D. Simon$^{2}$,
Zachary Slepian$^{23}$,
Marcelle Soares-Santos$^{15}$,
Greg Tarl{\'e}$^{11}$,
Ian Thompson$^{2}$,
Monica Valluri$^{11}$,
Risa H. Wechsler$^{25,26}$,
Martin White$^{6,13}$,
Michael J. Wilson$^{1}$,
Christophe Y\`eche$^{9}$,
Dennis Zaritsky$^{6}$

\newpage

$^{1}$ \LBL \\
$^{2}$ \Carnegie \\
$^{3}$ \UCBSSL \\
$^{4}$ \EPFLEng \\
$^{5}$ \Tsinghua \\
$^{6}$ \UAS \\
$^{7}$ \UNIPD \\
$^{8}$ \EPFL \\
$^{9}$ \IRFU \\
$^{10}$ \MPHeidelberg \\
$^{11}$ \Umich \\
$^{12}$ \UCBP \\
$^{13}$ \UCB \\
$^{14}$ \Pitt \\
$^{15}$ \Brandeis \\
$^{16}$ \PI \\
$^{17}$ \OU \\
$^{18}$ \NYU \\
$^{19}$ \OSU \\
$^{20}$ \NOAO \\
$^{21}$ \Wyoming \\
$^{22}$ \UCSC \\
$^{23}$ \UFL \\
$^{24}$ \UCL \\
$^{25}$ \Stanford \\
$^{26}$ \SLAC \\
$^{27}$ \Siena \\
$^{28}$ \KSU \\
$^{29}$ \FNAL \\
$^{30}$ \UR \\
$^{31}$ \Yale \\

\newpage

\justify

\section{Abstract}
MegaMapper is a proposed ground-based experiment to measure Inflation parameters and Dark Energy from galaxy redshifts at 2<z<5. A 6.5-m Magellan telescope will be coupled with DESI spectrographs to achieve multiplexing of 20,000. MegaMapper would be located at Las Campanas Observatory to fully access LSST imaging for target selection.

\section{Key Science Goals and Objectives} 
\subsection{Cosmic Origins}
The mechanisms driving the accelerated expansion of the Universe in its very first moments (Inflation) and at late times (Dark Energy) represent some of the most important open problems in fundamental physics, and have been the subject of several of the Astro2020 Science White Papers.

Primordial non-Gaussianity \cite{Meerburg:2019qqi} has been identified as one of the most powerful tools to study Inflation and the origin of the primordial fluctuations. While the simplest inflationary models predict Gaussian initial conditions, large classes of models predict  deviations from Gaussianity that leave specific imprints on the galaxy power spectrum and bispectrum \cite{Meerburg:2019qqi, Alvarez:2014vva}, thus probing physics at scales inaccessible to Earth-based colliders. Distinguishing multi-field from single-field inflation requires bounds on the ``local'' non-Gaussianity parameter to be considerably better than $\sigma(f_{NL}^{\rm local}) \lesssim 1$, or about an order of magnitude improvement over current constraints \cite{Akrami:2019izv}.

Our understanding of Dark Energy \cite{Slosar:2019flp} will greatly benefit from measuring the expansion and growth of fluctuations throughout cosmic history. A lever arm that extends from low redshift (Dark Energy era) to high redshifts (matter domination era) will tightly constrain large classes of theories, and test possible modifications to General Relativity.

The large volume available at $z > 2$ will allow us to increase the number of modes measured by over an order of magnitude over current experiments, opening up an uncharted territory with huge discovery potential.

In \cite{Ferraro:2019uce}, we have shown that a high-redshift spectroscopic survey covering the range $2 < z < 5$ can radically improve constraints on Inflation and Dark Energy, while at the same time relaxing assumptions such as a power-law primordial power spectrum \cite{Slosar:2019gvt, dePutter:2014hza}.  LSST imaging will allow us to select $\approx 100$ Million spectroscopic targets that span this redshift range. These targets are a combination of Lyman-Break galaxies (LBGs), selected using dropout techniques, and Lyman$-\alpha$ emitters (LAEs) for which broad-band color selection is possible \cite{Wilson:2019brt}.

In this white paper we propose a cost-effective version of such a survey, capable of achieving the stated science goals using proven technology. Simultaneously, this facility would confer a tremendous opportunity for ancillary science explorations that help answer a broad array of astronomical questions from the formation of the Milky Way to the evolution and growth of galaxies.
The MegaMapper would be located at Las Campanas Observatory in the southern hemisphere, and would have full access to LSST imaging for target selection.

\begin{figure}[h]
\includegraphics[width=16cm]{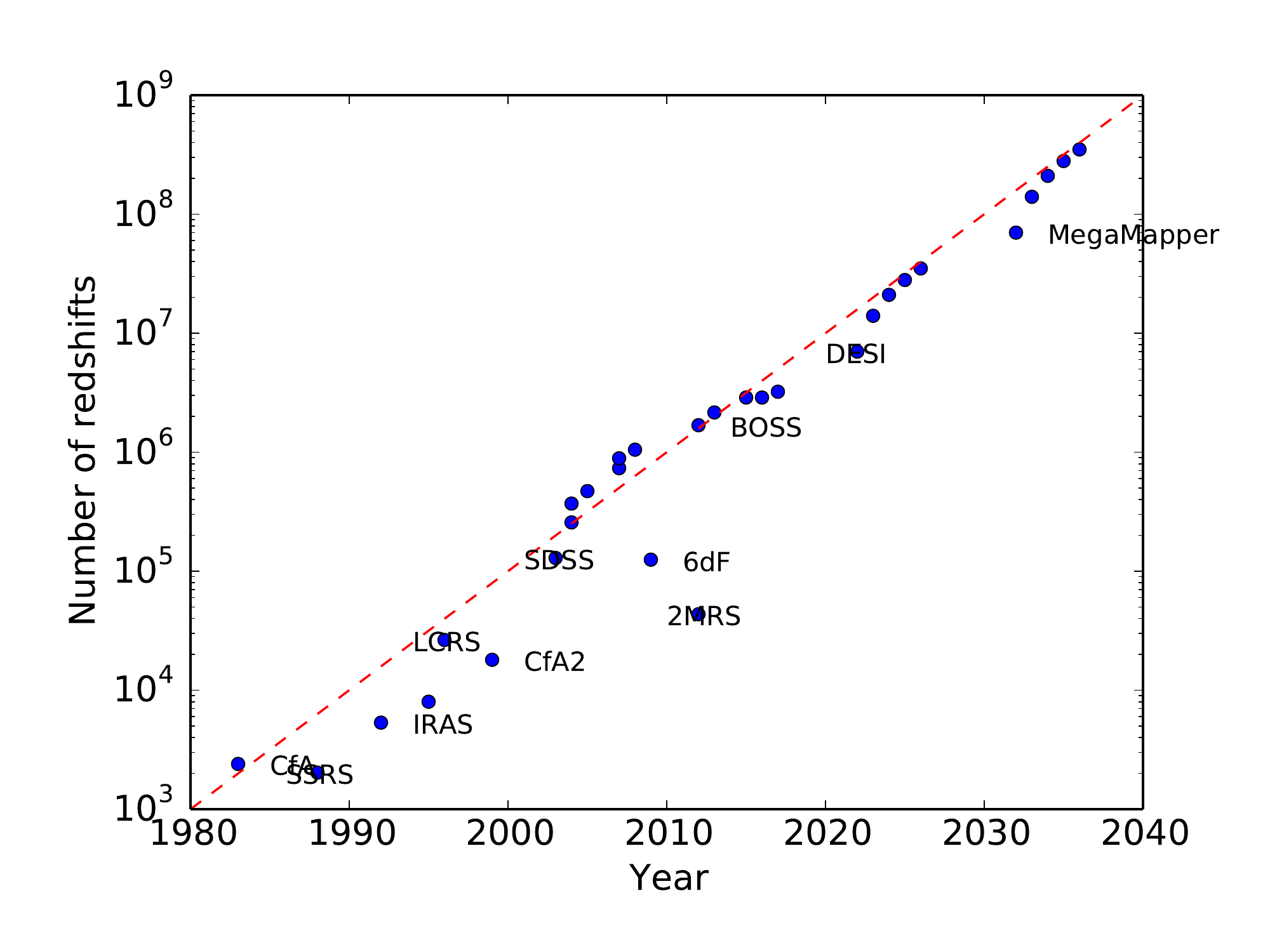}
\caption{Number of galaxy redshifts as a function of time for the largest cosmology surveys. The dotted line represents an increase of survey size by a factor of 10 every decade. Fielding the MegaMapper in ten years maintains this pace into the 2030s, and enables the Inflation and Dark Energy measures proposed in this and other white papers.}
\label{/3.ig:spectro_history}
\end{figure}

\subsection{Cosmology Science Forecasts: Inflation and Dark Energy}
In \cite{Ferraro:2019uce} we have identified two samples, a more optimistic ``idealized'' sample based on the LSST target density of LAEs and LBGs, and a ``fiducial'' sample, based on conservative redshift success rates and assumptions about line strengths (see \cite{Ferraro:2019uce} for the sample specifications and \cite{Wilson:2019brt} for background about selection and sample properties). 

We have found that both samples can significantly cross the theoretical threshold $\sigma(f_{NL}^{\rm local}) \lesssim 1$, surpassing the current and future CMB bounds by an order of magnitude. In particular, we find $\sigma(f_{NL}^{\rm local}) = 0.11$ and 0.073 for the fiducial and idealized samples respectively when including both the power spectrum and bispectrum.
Improvements by a factor of two or larger over the current bounds are also expected for the equilateral and orthogonal shapes \cite{Ferraro:2019uce}. Particular care should be taken in the telescope and survey design to control systematics to allow such precise measurements of primordial non-Gaussianity.

Using a combination of Redshift-Space Distortions (RSD) and Baryon Acoustic Oscillations (BAO), we can measure the fraction of Dark Energy $\Omega_{DE}$ to better than 1\% all the way to $z \approx 4.5$ for the ``idealized'' sample, and better than 2\% up to $z \approx 5$ for the ``fiducial'' sample (see Figure 1 in \cite{Ferraro:2019uce}). Other notable improvements include a factor of two better determination of the spatial Curvature (compared to DESI + Planck), and a factor of $\gtrsim 2.5$ improvement for the Dark Energy figure of merit (Table 4 in \cite{Ferraro:2019uce}).

\begin{table}[h!]
\caption{Survey speeds for multi-fiber spectrographs as measured by the product of the telescope clear aperture, number of fibers and losses from mirror reflections.  This speed assumes a dedicated program, which would not be possible in all cases.  Keck/FOBOS\cite{2019arXiv190707195B}, MSE\cite{2019arXiv190707192M}, SpecTel\cite{2019arXiv190706797E} and MegaMapper are proposed experiments.  LSSTspec\cite{2019arXiv190504669S} is a notional number using MegaMapper positioners on the LSST focal plane, if optical design limitations could be overcome injecting f/1.2 light into fibers.}
\label{tab_surveyspeed}
\begin{tabular}{lrrrrr@{.}l}\hline
Instrument (year)     & Primary/m$^2$ &  Nfiber  &  Reflections&  Product  & Speed vs& \ SDSS \\
          \hline
SDSS (1999)      &   3.68      &     640  &   0.9$^2$     &     1908  &    1&00 \\
BOSS (2009)     &   3.68      &    1000  &   0.9$^2$     &     2980  &    1&56 \\
DESI (2019)     &   9.5       &    5000  &   0.9$^1$     &   42,750  &   22&4 \\
PFS  (2020)     &  50         &    2400  &   0.9$^1$     &  108,000  &   56&6 \\
4MOST (2022)    &  12         &    1624  &   0.9$^2$     &      15,800  &    8&3 \\
{\bf MegaMapper}&    {\bf 28}       &    {\bf 20,000}&     {\bf 0.9}$^{\bf 2}$     &  {\bf 454,000}  &  {\bf 238}&  \\
Keck/FOBOS & 77.9       &    1800  &   0.9$^3$     &  102,000  &  53&6 \\
MSE       &  78         &    3249  &   0.9$^1$     &  228,000  &  119& \\
LSSTspec  &  35.3       &    8640  &   0.9$^3$     &  222,000  &  116& \\
SpecTel   &  87.9       &   15,000 &   0.9$^2$     & 1,070,000  &  560& \\
\hline
\end{tabular}
\end{table}

\begin{table}[h!]
\caption{Survey speeds as measured by the raw product of collecting area and field-of-view.  This is the appropriate metric for a wide-area survey with sparse targets.  Even without taking full advantage of multiplexing, the MegaMapper survey speed is competitive with larger telescopes owing to its large field-of-view.}
\label{tab_surveyarea}
\begin{tabular}{lrrrrr@{.}l}\hline
Instrument (year)      &    Primary/m$^2$  & FOV/deg$^2$ & Reflections & Product &  Speed vs& \ SDSS \\
\hline
SDSS (1999)   &     3.68     &    7.06      &  0.9$^2$   &      21.0    &    1&00 \\
BOSS (2009)   &     3.68     &    7.06      &  0.9$^2$   &      21.0    &    1&00 \\
DESI (2019)  &     9.5    &      8.04   &   0.9$^1$    &     68.7    &    3&27 \\
PFS (2020)   &    50      &      1.33   &   0.9$^1$    &      59.9    &    2&85 \\
4MOST (2022) &    12      &      4.90   &   0.9$^2$    &    58.8    &    2&80 \\
{\bf MegaMapper} &  {\bf 28}     &      {\bf 7.06}   &   {\bf 0.9}$^{\bf 2}$   &     {\bf 160.} &       {\bf 7}&{\bf{62}} \\
Keck/FOBOS  & 77.9   &     0.087  &   0.9$^3$    &     4.94    &   0&23 \\
MSE    &    78       &     1.52   &   0.9$^1$   &     107.     &    5&10 \\
LSSTspec   &    35.3    &      9.60   &   0.9$^3$   &     247.  &      11&76 \\
SpecTel    &    87.9    &      4.91   &   0.9$^2$   &     350.  &      16&65 \\
\hline
\end{tabular}
\end{table}

\subsection{Additional Science Opportunities}
The facility constructed in pursuit of the Cosmic Origins program described above also serves a broad range of additional astrophysical and cosmological objectives.  Some of these would be addressed coincident with the 5-year cosmology key project, while others could be pursued after completion of that project.

A wide-field spectroscopic survey in the southern hemisphere will greatly enhance the LSST science returns, as identified by several Astro 2020 science white papers \cite{Mandelbaum:2019zej, Newman:2019doi, Bechtol:2019acd}. Calibration of photometric redshifts is possible over the whole range of LSST sources through cross-correlation techniques with the spectroscopic sample. The large overlap area will enable a reduction in the statistical errors to meet the stringent LSST requirements \cite{Mandelbaum:2019zej}.  The availability of a large overlapping spectroscopic sample will allow cross correlation of these galaxies with the faint LSST sources to better constrain the Intrinsic Alignment models.
Moreover, a combination of lensing amplitude provided by LSST, together with growth measurements through RSD can provide a powerful test of General Relativity on cosmological scales.
Finally it can provide redshifts for strong gravitational lenses and type Ia supernovae discovered in LSST, allowing their cosmological interpretation.

It will also be a key complement to the astrophysical studies of Dark Matter \cite{Bechtol:2019acd, Drlica-Wagner:2019xan}: by measuring the velocity dispersion of faint Milky Way satellites, the mass and density can be inferred and compared to theoretical predictions, in an environment where baryon effects are expected to be minimal. Similarly, the perturbations and gaps in the Milky Way stellar streams created by encounters with Dark Matter substructure create a characteristic velocity signal that can be measured with spectroscopy to help determine the perturber mass \cite{Drlica-Wagner:2019xan}. Moreover, it can aid the study of merger dynamics in galaxy clusters, and provide constraints on self-interacting Dark Matter.

Additionally, it can allow the study of the environment dependence of galaxy evolution to $z \sim 2$ and stellar population and kinematics and dense tomography of the intergalactic medium.  While it is beyond the scope of what is presented here, future instrumentation on such a facility in the post-survey period could enable a tremendous range of science from planet to galaxy formation and evolution.

\section{Technical Overview}

The overall design seeks to optimize survey speed with a judicious choice of telescope aperture and instrument multiplexing.  A new set of optical models,  presented for the first time in this white paper, achieves a speed (as measured by $A \cdot \Omega$) that is difficult to match with either smaller or larger telescopes (see Table \ref{tab_surveyspeed}).  The overall telescope design is nearly identical to the existing Magellan 6.5-m telescopes, although the preferred design modifies the central hole to be larger.  The instrument achieves a multiplexing of 20,000 with 32 identical focal plane petals with 625 fiber positioners feeding one spectrograph.  16 of these spectrographs already exist or are in construction for the DESI and SDSS-V projects (see \cite{2016arXiv161100037D}), with another 16 spectrographs required for this full MegaMapper instrument.

\noindent
{\bf Telescope}:
The telescope concept is based on the highly successful Magellan telescope design.  This constitutes a lightweight, honeycomb structure, 6.5~m borosilicate glass primary mirror built by the University of Arizona Mirror Lab. A baseline design is considered, which adopts the same optical prescription as Magellan I and II (i.e. a f/1.25 paraboloid), equipped with a 2.4 m hyperbolic secondary mirror ($\sim$70\% larger in diameter than the current f/11 Gregorian secondaries), and a 5-lens wide field corrector that provides a 3.0~deg diameter field-of-view on a Cassegrain focal plane fed at f/3.6 (see Figure \ref{fig:opticaldesign2}). An ADC is designed as part of the corrector. The first and largest element in the corrector is 1.8~m in diameter, with the other lenses also being meter sized \footnote{The LSST project had Arizona Optical Systems build and coat a 1.5m lens for their system.}. The large secondary and sky baffles imply a central obscuration of $\sim$20\% of the area of the primary mirror, significantly larger than the $\sim5$\% obscuration in the current Magellans but still less than either SDSS or DESI. This baseline design requires a larger central hole in the primary mirror than the 1.3~m hole in Magellan I and II. This would require casting the mirror with a custom mold at the Mirror Lab.

The f/3.6 telecentric focal plane has a diameter of 1230~mm, which at a plate scale of 0.113~mm/" corresponds to a 3.0~deg diameter FOV. Figure \ref{fig:spot} presents spot diagrams at different zenith angles for the telescope's preliminary optical design described above, with $<23$~$\mu$m rms radius ($\sim 0.4$" FWHM on sky) across the full FOV, which has a maximum of 2.7\% vignetting at the field edge. This is well matched to the median seeing at LCO (FWHM$\simeq$0.6") and the typical fiber size used by replicable fiber-fed spectrographs. At this platescale a 107~$\mu$m optical fiber (identical to the DESI fibers) subtends a diameter of 0.94" on the sky, which is near the optimal fiber size for point sources or compact high-redshift galaxies in the sky-noise limit. 

\begin{figure}[t]
\includegraphics[width=12cm]{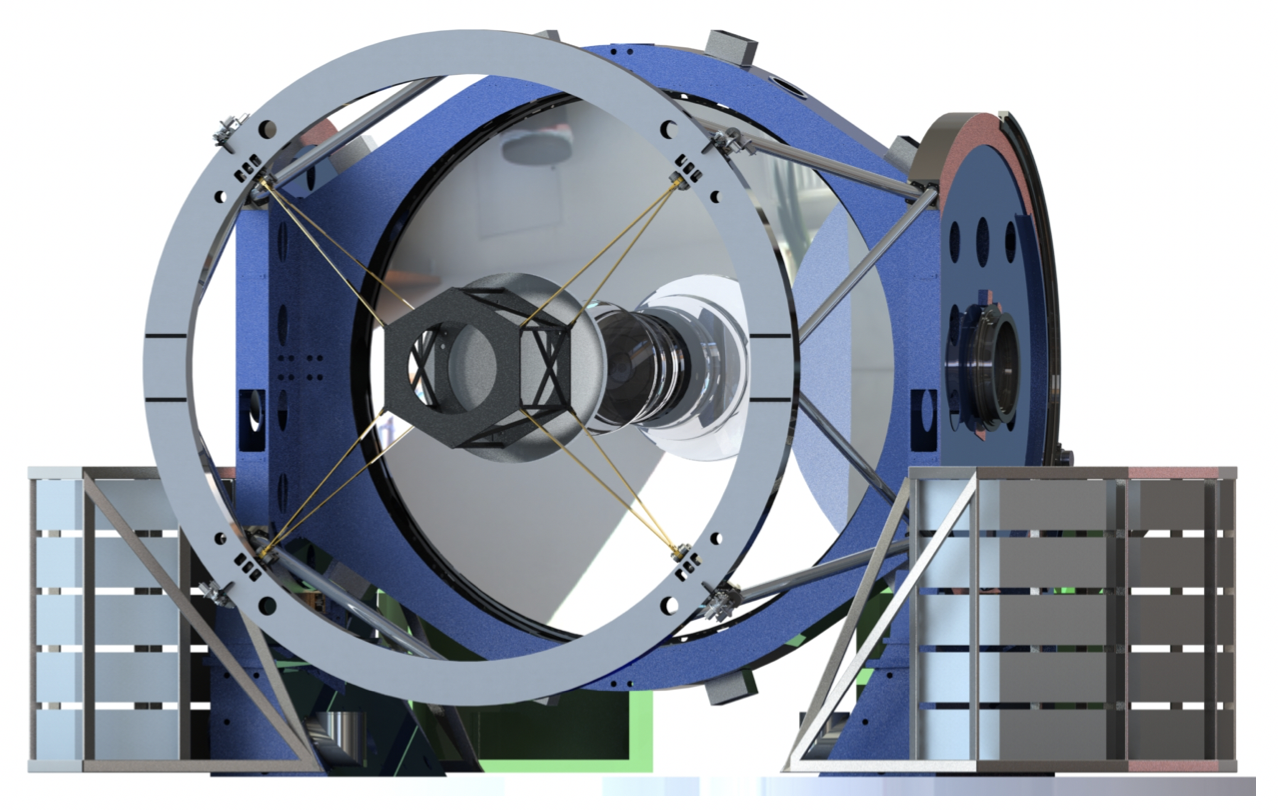}
\caption{Rendering of the Magellan-style telescope with the secondary mirror and 7-element corrector, pointed towards the horizon. The 32 MegaMapper spectrographs are parked on the base  with a fiber run that is substantially shorter than the 50-meter run for DESI.
}
\label{fig:jeffscad}
\end{figure}

\begin{figure}[h]
\includegraphics[width=12cm]{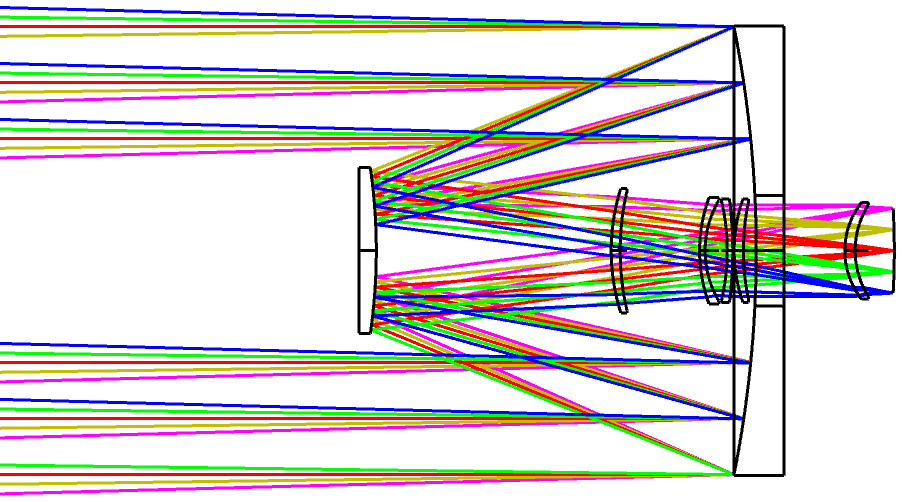}
\caption{The optical systems on the Magellan telescope.}
\label{fig:opticaldesign2}
\end{figure}

\begin{figure}[h]
\includegraphics[width=10cm]{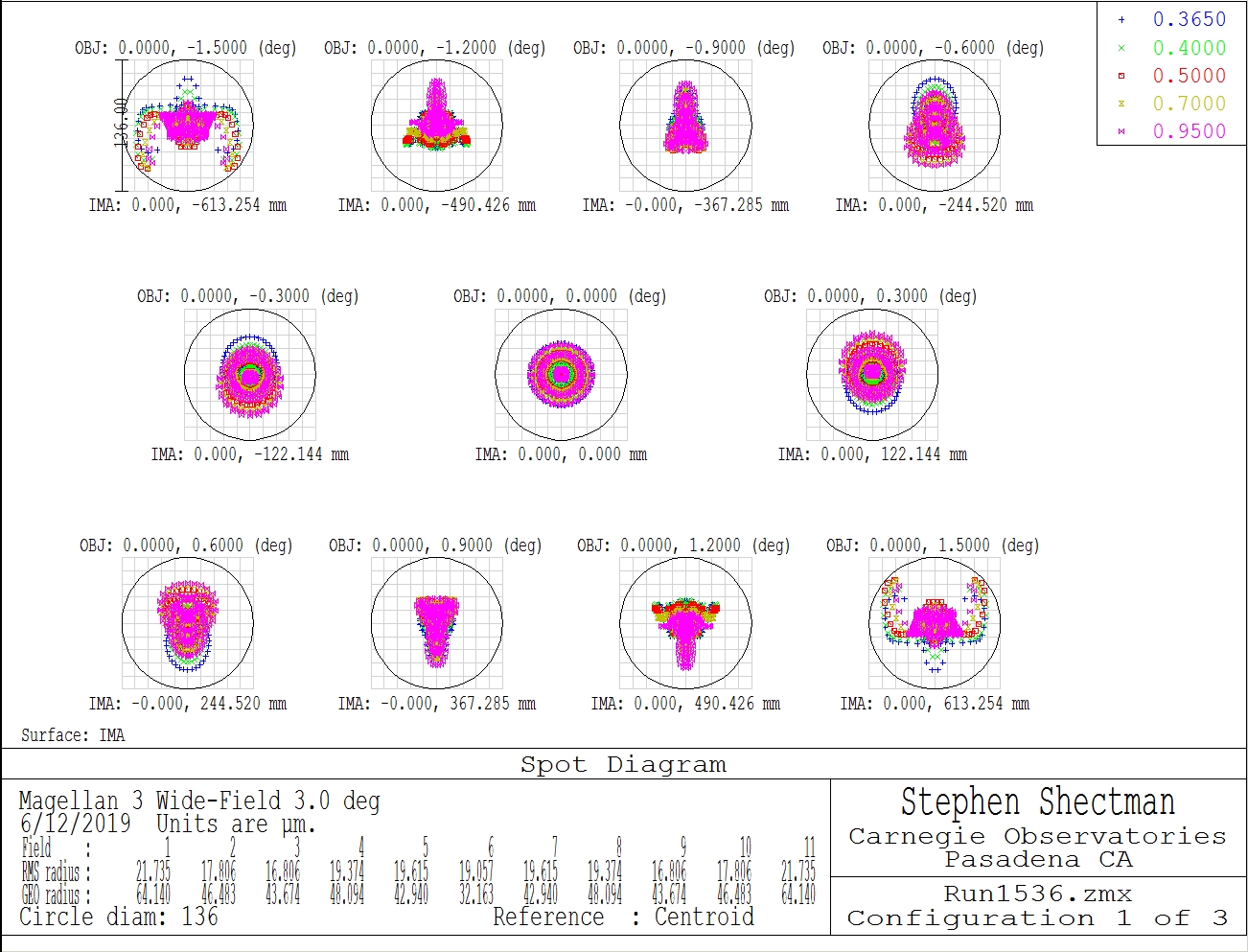}
\caption{Spot diagram for the telescope and corrector optics.}
\label{fig:spot}
\end{figure}

The telescope mount, enclosure, and auxiliary facilities could be identical to the current Magellan design. However, particular sub-systems should be redesigned and updated.  Figure \ref{fig:jeffscad} shows the MegaMapper in the Magellan CAD.  Some subsystems would benefit from advances of technology since the Magellan 1 \& 2 builds, in particular for the sensor and control systems.

Two sites at Las Campanas Observatory are exceptional choices for the MegaMapper. Cerro Alcaino and Manqui's Ridge were each extensively tested as potential sites for the Giant Magellan Telescope. The sites are well-characterized in terms of image quality and environmental statistics, and offer excellent observing conditions for a new large telescope. 

{\bf Focal Plane}:  The focal plane is physically large enough to accommodate 20,000 zonal fiber positioners with a  center-to-center pitch of 6.5 mm.  The focal plane will be divided into 32 segments, where each segment is constructed as a complete instrument with 625 fibers feeding one spectrograph.  Such a segmentation of the instrument has proven critical to efficient construction, integration and testing of highly-multiplexed spectrographs such as MUSE, VIRUS and DESI.

Each individual fiber positioner is composed of two precision, mechanical gearmotors.  Such a ``theta-phi'' motion is the basis of both the DESI and SDSS-V positioner design.  These have the benefit of fast positioning time, high accuracy, and maintaining optical telecentricity.  DESI has completed their focal plane of 5000 positioners, demonstrating that such a system can be mass produced, reliably controlled, positioned to an accuracy of $<2$~micron, and reconfigured in a time envelope of less than one minute.

MegaMapper represents an evolution of the DESI focal plane with smaller positioners and other design modifications based upon our experiences building the DESI and SDSS-V focal planes
to facilitate mass production, integration and testing.     

{\bf Spectrographs}:  The spectrographs would be identical to those under construction for DESI and SDSS-V. These spectrographs went through extensive design studies, and are optimized to measure redshifts of faint targets in the sky-noise limit.  The performance of these spectrographs have been shown to exceed their design goals in delivered optical quality and throughput.  We would choose to somewhat increase the number of fibers feeding each spectrograph from 500 to 625 by decreasing the fiber spacing at the spectrograph slit.  This is supported by the as-delivered spot quality in the spectrographs.

Each spectrograph is fed by a pseudo-slit with 625 fibers, with dichroics dividing collimated light into three cameras.  Each camera has gratings, optics and CCDs that are optimized for its wavelength range in 360--555, 555--656, and 656--980 nm channels.  The spectral resolution runs from 2000 on the blue end (at 360~nm) increasing to a resolution of 5500 on the red end (at 980~nm) in order to work between bright sky lines.  The as-built efficiences are 70--90\% across the full optical range.

\begin{figure}[h]
\includegraphics[width=16cm]{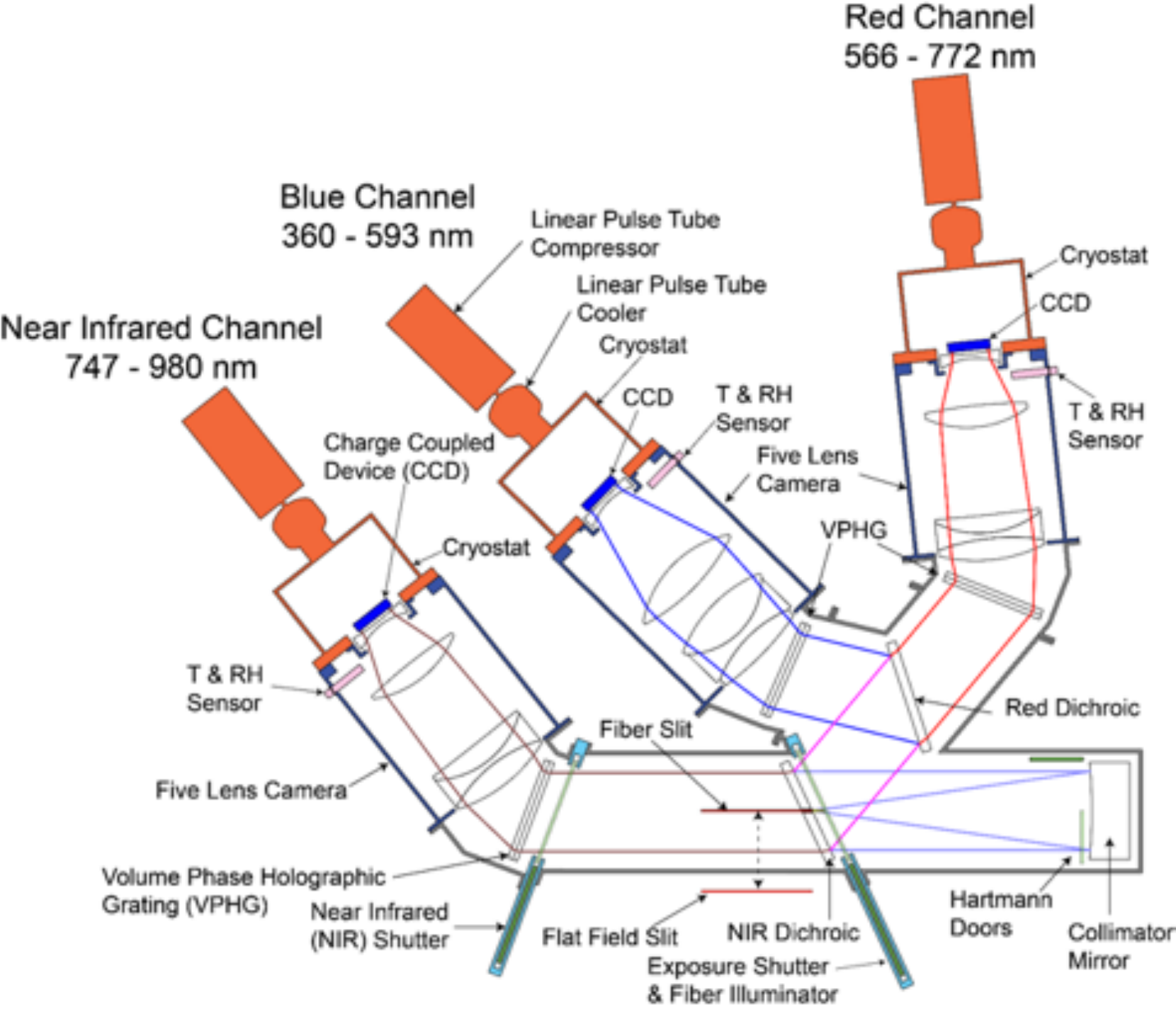}
\caption{Optical model of the DESI and SDSS-V spectrographs, which are identical systems aside from the CCD packages. As-built efficiencies are 70--90\% across the full optical spectrum.}
\label{fig:spectro_model}
\end{figure}

{\bf Data System}: The MegaMapper data systems will be a continuation of the data system as developed by the DESI team and operated on the NERSC high-performance computing platform.  The DESI data reduction from the raw pixel level to fully-calibrated spectra and redshift-fitting is state-of-the-art today, and will be maintained and updated at least through the DESI key project from 2020-2025.  Poisson-limited spectra for faint targets is achieved through
a combination of
stability of the spectrographs, stability of the PSF with theta-phi positioners, and a rigorous forward-modeling of the spectral extraction and sky-subtraction \cite{2010PASP..122..248B}.

The software modifications that will be required for MegaMapper are a training of spectroscopic templates based upon commissioning data. 

\section{Technology Drivers}

The MegaMapper is designed to primarily recycle and replicate existing, cost-effective technologies.  Technology development is required for the smaller-pitch theta-phi fiber positioners.  An R\&D program for this effort has been proposed through the DOE Cosmic Visions process \cite{2018arXiv180207216D}.

\section{Organization, Partnerships, and Current Status}

The MegaMapper concept is envisioned to be a partnership led by Carnegie Observatories and the Lawrence Berkeley National Laboratory (LBNL).  Carnegie would be responsible for the construction of the telescope and 5-lens optical corrector. LBNL would be responsible for the instrument, including the robotic focal plane and spectrographs.  Telescope operations would be managed by Carnegie, and instrument operations and data reduction by LBNL.  This partnership would continue through the operations phase of a 5-year cosmology key project.

\section{Schedule}

The notional schedule below reproduces the Magellan 2 (Clay) telescope construction schedule (offset by 25 years) and the DESI instrument schedule (offset by 10 years).  The instrument schedule is denoted by DOE critical decision milestones that correspond to a conceptual design (CD-1), preliminary design (CD-2) and final design (CD-3).

\begin{table}[h]
%\caption{}
\label{tab_milestones}
\begin{tabular}{ll}\hline
Year & Milestone \\
\hline
2021 & Telescope optical design finalized \\
2022 & Project approval, DOE Critical Decision 0 (Mission Need) \\
2023 & Casting of mirrors and begin aquisition of corrector blanks \\
2025 & DOE Critical Decision 1 (Approve Alternative Selection) \\
2026 & DOE Critical Decision 2 (Approve Alternative Analysis and Cost Range) \\
2027 & DOE Critical Decision 3 (Approve Start of Construction) \\
2027 & Telescope first light \\
2028 & Focal plane and spectrograph installation start \\
2029 & Operations start \\
\hline
\end{tabular}
\end{table}

\section{Cost Estimates}

A detailed costing estimate for MegaMapper has not yet been performed.  However, the costs can be estimated from the two Magellan telescopes, the DESI corrector, and the identical DESI and SDSS-V spectrographs.

\noindent
{\bf Telescope facility}:  The telescope facility cost is defined as the telescope, primary mirror and support system, secondary mirror cage and position actuators, buildings, aluminizing chamber and utilities.  The costing has been estimated from the 
the inflation-adjusted cost of building the dual Magellan telescope facilities.  In 2020 dollars, the construction of BOTH of these telescopes is approximately {\$}100~M, and we estimate {\$}70~M for one telescope.

{\bf Corrector}:  The 5-element corrector is similar in complexity to the DESI 6-element corrector, with 2X more mass in the glass and in the mechanical structure.  The elements for MegaMapper are larger, but have the advantage of being all fused silica lenses.
 
 {\bf Focal plane}:  The cost estimate for the 20,000-fiber MegaMapper focal plane is \$10~M.  This would be a factor of 2 reduction of cost per-positioner as compared to the 5000-fiber DESI focal plane.  This cost reduction would come from the motor part costs and the QA testing costs, which were the dominant costs for the DESI focal plane.  Although the fiber pitch is smaller for MegaMapper, the motors will be larger (5~mm) with somewhat loosened accuracy specifications.  The QA testing will be mitigated by having the motor vendors construct a more integrated package with fewer total parts, which was the approach successfully adopted for the SDSS-V construction project.
 
 {\bf Fiber system}: The fiber system will consist of 32 separate bundles of 625 fibers, with each bundle connecting one segment of the focal plane with one spectrograph.  Our team has extensive experience building up to 9000 fibers over several generations of SDSS instruments, and 5000 fibers for the DESI instrument.  Preliminary design will trade between monolithic fiber runs (as done for SDSS-I, SDSS/BOSS), connectorized fibers (as done for SDSS/APOGEE), or plasma-fused fibers (as done for DESI).  The detailed choices for MegaMapper will likely be determined by integration and installation considerations. 

{\bf Spectrographs}:  MegaMapper is designed to use the 10 DESI and 6 SDSS-V spectrographs without modification, after they have completed their key projects plus an additional three-year run time to 2028.   Those spectrographs would account for 50\% of the instrument, thus requiring only 16 additional spectrographs to complete the instrument.  At a unit cost of \$1.5~M per spectrograph, the cost is \$24~M.

\begin{table}[htbp]
\caption{MegaMapper costing and basis of estimates}
\label{tab_cost}
\begin{tabular}{lrl}\hline
Component & Cost (\$M) & Basis of estimate \\
\hline
Telescope facility & 70 & Magellan 1 \& 2 \\
Secondary & 10 & Vendor ROM \\
Corrector & 20 & DESI \\
Focal plane & 10 & DESI \\
Fibers    &  5 & DESI \\
Spectrographs & 24 & DESI and SDSS-V \\
\hline
Total  & 139 & \\
\hline
\end{tabular}
\end{table}

\section{Conclusions}

The MegaMapper is designed to be a cost-effective approach to achieving a  survey speed 12X faster than the Dark Energy Spectroscopic Instrument (DESI).  The 6.5-m primary mirror has 3X the collecting area of DESI, yet is no more challenging in the optical design for the focal plane and the spectrographs.  The telescope costing is based upon the Magellan 6.5-m telescopes, and the instrument costing is based upon the DESI experiment. 
Our approach thus harnesses the power of efficient mid-scale facility construction (i.e. the Magellan telescopes) and survey design and operation (i.e. the SDSS/DESI surveys).
In combination with the next-generation of CMB experiments, MegaMapper promises to explore the full range of single-field and multi-field Inflation models.

\newpage
\bibliographystyle{utphys}
\bibliography{main}

\providecommand{\href}[2]{#2}\begingroup\raggedright\begin{thebibliography}{10}

\bibitem{Meerburg:2019qqi}
P.~D. Meerburg {\em et~al.}, ``{Primordial Non-Gaussianity},''
\href{http://arxiv.org/abs/1903.04409}{{\ttfamily arXiv:1903.04409
  [astro-ph.CO]}}.
%%CITATION = ARXIV:1903.04409;%%.

\bibitem{Alvarez:2014vva}
M.~Alvarez {\em et~al.}, ``{Testing Inflation with Large Scale Structure:
  Connecting Hopes with Reality},''
\href{http://arxiv.org/abs/1412.4671}{{\ttfamily arXiv:1412.4671
  [astro-ph.CO]}}.
%%CITATION = ARXIV:1412.4671;%%.

\bibitem{Akrami:2019izv}
{\bfseries Planck} Collaboration, Y.~Akrami {\em et~al.}, ``{Planck 2018
  results. IX. Constraints on primordial non-Gaussianity},''
\href{http://arxiv.org/abs/1905.05697}{{\ttfamily arXiv:1905.05697
  [astro-ph.CO]}}.
%%CITATION = ARXIV:1905.05697;%%.

\bibitem{Slosar:2019flp}
A.~Slosar {\em et~al.}, ``{Dark Energy and Modified Gravity},''
\href{http://arxiv.org/abs/1903.12016}{{\ttfamily arXiv:1903.12016
  [astro-ph.CO]}}.
%%CITATION = ARXIV:1903.12016;%%.

\bibitem{Ferraro:2019uce}
S.~Ferraro {\em et~al.}, ``{Inflation and Dark Energy from spectroscopy at $z >
  2$},''
\href{http://arxiv.org/abs/1903.09208}{{\ttfamily arXiv:1903.09208
  [astro-ph.CO]}}.
%%CITATION = ARXIV:1903.09208;%%.

\bibitem{Slosar:2019gvt}
A.~Slosar {\em et~al.}, ``{Scratches from the Past: Inflationary Archaeology
  through Features in the Power Spectrum of Primordial Fluctuations},''
\href{http://arxiv.org/abs/1903.09883}{{\ttfamily arXiv:1903.09883
  [astro-ph.CO]}}.
%%CITATION = ARXIV:1903.09883;%%.

\bibitem{dePutter:2014hza}
R.~de~Putter, E.~V. Linder, and A.~Mishra, ``{Inflationary Freedom and
  Cosmological Neutrino Constraints},''
  \href{http://dx.doi.org/10.1103/PhysRevD.89.103502}{{\em Phys. Rev.}
  {\bfseries D89} no.~10, (2014) 103502},
\href{http://arxiv.org/abs/1401.7022}{{\ttfamily arXiv:1401.7022
  [astro-ph.CO]}}.
%%CITATION = ARXIV:1401.7022;%%.

\bibitem{Wilson:2019brt}
M.~J. Wilson and M.~White, ``{Cosmology with dropout selection: Straw-man
  surveys and CMB lensing},''
\href{http://arxiv.org/abs/1904.13378}{{\ttfamily arXiv:1904.13378
  [astro-ph.CO]}}.
%%CITATION = ARXIV:1904.13378;%%.

\bibitem{2019arXiv190707195B}
K.~{Bundy}, K.~{Westfall}, N.~{MacDonald}, R.~{Kupke}, M.~{Savage},
  C.~{Poppett}, A.~{Alabi}, G.~{Becker}, J.~{Burchett}, P.~{Capak}, A.~{Coil},
  M.~{Cooper}, D.~{Cowley}, W.~{Deich}, D.~{Dillon}, J.~{Edelstein},
  P.~{Guhathakurta}, J.~{Hennawi}, M.~{Kassis}, K.~G. {Lee}, D.~{Masters},
  T.~{Miller}, J.~{Newman}, J.~{O'Meara}, J.~X. {Prochaska}, M.~{Rau},
  J.~{Rhodes}, R.~M. {Rich}, C.~{Rockosi}, A.~{Romanowsky}, C.~{Schafer},
  D.~{Schlegel}, A.~{Shapley}, B.~{Siana}, Y.~S. {Ting}, D.~{Weisz},
  M.~{White}, B.~{Williams}, G.~{Wilson}, M.~{Wilson}, and R.~{Yan}, ``{FOBOS:
  A Next-Generation Spectroscopic Facility at the W. M. Keck Observatory},''
  {\em arXiv e-prints} (Jul, 2019) arXiv:1907.07195,
  \href{http://arxiv.org/abs/1907.07195}{{\ttfamily arXiv:1907.07195
  [astro-ph.IM]}}.

\bibitem{2019arXiv190707192M}
J.~{Marshall}, J.~{Bullock}, A.~{Burgasser}, K.~{Chambers}, D.~{DePoy},
  A.~{Dey}, N.~{Flagey}, A.~{Hill}, L.~{Hillenbrand}, D.~{Huber}, T.~{Li},
  S.~{Juneau}, M.~{Kaplinghat}, M.~{Mateo}, A.~{McConnachie}, J.~{Newman},
  A.~{Petric}, D.~{Schlegel}, A.~{Sheinis}, Y.~{Shen}, D.~{Simons},
  M.~{Strauss}, K.~{Szeto}, K.-V. {Tran}, C.~{Y{\`e}che}, and {the MSE Science
  Team}, ``{The Maunakea Spectroscopic Explorer},'' {\em arXiv e-prints} (Jul,
  2019) arXiv:1907.07192, \href{http://arxiv.org/abs/1907.07192}{{\ttfamily
  arXiv:1907.07192 [astro-ph.IM]}}.

\bibitem{2019arXiv190706797E}
R.~{Ellis}, K.~{Dawson}, J.~{Bland-Hawthorn}, R.~{Bacon}, A.~{Bolton},
  M.~{Bremer}, J.~{Brinchmann}, K.~{Bundy}, C.~{Conroy}, B.~{Delabre},
  A.~{Dey}, A.~{Drlica-Wagner}, J.~{Greene}, L.~{Guzzo}, J.~{Johnson},
  A.~{Leauthaud}, K.-G. {Lee}, L.~{Pasquini}, L.~{Pentericci}, J.~{Richard},
  H.-W. {Rix}, C.~{Rockosi}, D.~{Schlegel}, A.~{Slosar}, M.~{Strauss},
  M.~{Takada}, E.~{Tolstoy}, and D.~{Watson}, ``{SpecTel: A 10-12 meter class
  Spectroscopic Survey Telescope},'' {\em arXiv e-prints} (Jul, 2019)
  arXiv:1907.06797, \href{http://arxiv.org/abs/1907.06797}{{\ttfamily
  arXiv:1907.06797 [astro-ph.IM]}}.

\bibitem{2019arXiv190504669S}
C.~W. {Stubbs} and K.~{Heitmann}, ``{Report on LSST Next-generation
  Instrumentation Workshop, April 11, 12 2019},'' {\em arXiv e-prints} (May,
  2019) arXiv:1905.04669, \href{http://arxiv.org/abs/1905.04669}{{\ttfamily
  arXiv:1905.04669 [astro-ph.IM]}}.

\bibitem{Mandelbaum:2019zej}
{\bfseries LSST Dark Energy Science} Collaboration, R.~Mandelbaum {\em et~al.},
  ``{Wide-field Multi-object Spectroscopy to Enhance Dark Energy Science from
  LSST},''
\href{http://arxiv.org/abs/1903.09323}{{\ttfamily arXiv:1903.09323
  [astro-ph.CO]}}.
%%CITATION = ARXIV:1903.09323;%%.

\bibitem{Newman:2019doi}
{\bfseries LSST Dark Energy Science} Collaboration, J.~A. Newman {\em et~al.},
  ``{Deep Multi-object Spectroscopy to Enhance Dark Energy Science from
  LSST},''
\href{http://arxiv.org/abs/1903.09325}{{\ttfamily arXiv:1903.09325
  [astro-ph.CO]}}.
%%CITATION = ARXIV:1903.09325;%%.

\bibitem{Bechtol:2019acd}
K.~Bechtol {\em et~al.}, ``{Dark Matter Science in the Era of LSST},''
\href{http://arxiv.org/abs/1903.04425}{{\ttfamily arXiv:1903.04425
  [astro-ph.CO]}}.
%%CITATION = ARXIV:1903.04425;%%.

\bibitem{Drlica-Wagner:2019xan}
{\bfseries LSST Dark Matter Group} Collaboration, A.~Drlica-Wagner {\em
  et~al.}, ``{Probing the Fundamental Nature of Dark Matter with the Large
  Synoptic Survey Telescope},''
\href{http://arxiv.org/abs/1902.01055}{{\ttfamily arXiv:1902.01055
  [astro-ph.CO]}}.
%%CITATION = ARXIV:1902.01055;%%.

\bibitem{2016arXiv161100037D}
{DESI Collaboration}, A.~{Aghamousa}, J.~{Aguilar}, S.~{Ahlen}, S.~{Alam},
  L.~E. {Allen}, C.~{Allende Prieto}, J.~{Annis}, S.~{Bailey}, and
  C.~{Balland}, ``{The DESI Experiment Part II: Instrument Design},'' {\em
  arXiv e-prints} (Oct, 2016) arXiv:1611.00037,
  \href{http://arxiv.org/abs/1611.00037}{{\ttfamily arXiv:1611.00037
  [astro-ph.IM]}}.

\bibitem{2010PASP..122..248B}
A.~S. {Bolton} and D.~J. {Schlegel}, ``{Spectro-Perfectionism: An Algorithmic
  Framework for Photon Noise-Limited Extraction of Optical Fiber
  Spectroscopy},'' \href{http://dx.doi.org/10.1086/651008}{{\em PASP}
  {\bfseries 122} no.~888, (Feb, 2010) 248},
  \href{http://arxiv.org/abs/0911.2689}{{\ttfamily arXiv:0911.2689
  [astro-ph.IM]}}.

\bibitem{2018arXiv180207216D}
K.~{Dawson}, J.~{Frieman}, K.~{Heitmann}, B.~{Jain}, S.~{Kahn},
  R.~{Mandelbaum}, S.~{Perlmutter}, and A.~{Slosar}, ``{Cosmic Visions Dark
  Energy: Small Projects Portfolio},'' {\em arXiv e-prints} (Feb, 2018)
  arXiv:1802.07216, \href{http://arxiv.org/abs/1802.07216}{{\ttfamily
  arXiv:1802.07216 [astro-ph.CO]}}.

\end{thebibliography}\endgroup
\end{document}